\documentclass[12pt]{article}
\usepackage{epsf}
\newcommand{\beqn}{\begin{eqnarray}}
\newcommand{\eeqn}{\end{eqnarray}}

\newcommand{\Z}{{Z \!\!\! Z}}
\newcommand{\dd}{\mbox{d}}
\newcommand{\dD}{{\cal D}}
\newcommand{\cZ}{{\cal Z}}
\newcommand{\cO}{{\cal O}}

\newcommand{\cL}{{\cal L}}

\newcommand{\eq}[1]{(\ref{#1})}
\newcommand{\dual}[1]{{#1}^d}
\newcommand{\diff}{\partial}

\newcommand{\itep}{~\vspace{-1.5cm}\begin{flushright}
{\large ITEP-TH-10/99}\\
\vskip 0.3cm
{\large KANAZAWA 99-03}
\end{flushright}\vspace{1.0cm}}

\begin{document}
\begin{titlepage}
\itep
\begin{center}

{\large\bf Infrared Behaviour of the Gauge Boson Propagator} \vskip 0.2cm
{\large \bf in a Confining Theory}

\vskip 1.5cm
{\bf M.N.~Chernodub$^a$, M.I.~Polikarpov$^a$ and
V.I.~Zakharov$^{b}$}
\vskip 1.5cm
${}^a$  {\it Institute of Theoretical and Experimental Physics},\\
{\it B.Cheremushkinskaya 25, Moscow, 117259, Russia}
\vskip 0.5cm
\centerline{and}
\vskip 0.5cm
${}^b$ {\it Max-Planck Institut f\"ur Physics, Werner-Heisenberg
Institut,}\\ {\it 80805  Munich, Germany.}\\
\vskip 0.8cm
\end{center}

\begin{abstract}
We study the gauge boson propagator in the dual Abelian Higgs
theory which confines electric charges. The confinement is due to dual
Abrikosov-Nielsen-Olesen strings. We show that the infrared double
pole $1 \slash {(p^2)}^2$ in the propagator is absent {\it due to} the
confinement phenomenon.
Instead, specific angular singularities signal the eventual
set in of the confinement.
These angular singularities are a manifestation of
the confining strings.
\end{abstract}
\end{titlepage}

{\bf 1.} The most popular explanation of the quark confinement is the
dual superconductor model of QCD vacuum~\cite{tHMa76}. There is a lot
of numerical facts
demonstrating that in the abelian projection~\cite{tH} of the
lattice gluodynamics the monopoles are the relevant infrared degrees of
freedom responsible for the quark confinement and for formation of
confining string (see, $e.g.$, reviews~\cite{Reviews} and
references therein).

There exists yet another, so to say ``phenomenological''
point of view on the
quark confinement formulated in terms of the gluon propagator.
Namely, one argues \cite{West,BaBaZa88} that
if the gluon propagator has a double pole in the infrared
region then the quark--anti--quark potential gets
a linear attractive contribution
at large distances and this implies the quark confinement.
On the other hand, Gribov and
Zwanziger ~\cite{Zw91} argued that elimination of the gauge copies
leads to a counter-intuitive vanishing
of the gluon propagator in the infrared region. Some
analytical studies~\cite{alagribov} confirm this behaviour, while the
others~\cite{nogribov} predict a singular gluon propagator in the
infrared limit. Recent numerical simulations in lattice
gluodynamics~\cite{ZwNumerical} show that the
gluon propagator seems to
vanish at small momenta.
However, the systematic errors due to a finite lattice
size do not allow to reach a final conclusion.

In this paper we
study analytically the infrared behaviour of
the gauge boson propagator in a $U(1)$ theory with confinement.
Namely  we consider the dual Abelian Higgs theory in which magnetic
monopoles are condensed and electric charges are confined.
The motivation is that this theory
can be considered as an effective infrared theory of $SU(2)$
gluodynamics~\cite{MaSu89}. Moreover, it was found recently that the
vacuum of the lattice $SU(2)$ gluodynamics is well described by the dual
Abelian Higgs model in which monopoles play the role of the Higgs
particles \cite{effmod}. We define an analogue of the gluon propagator
in terms of abelian variables and show that the infrared double
pole in the propagator is absent in the confining phase
of the model.
We do not observe the vanishing
of the gluon propagator in the infrared region ~\cite{Zw91} either.
In the case considered,
the confinement properties of the theory are rather manifested through
string like singularities of the propagator. Since the dual Abelian Higgs
model imitates QCD in the infrared region a similar
behaviour might be exhibited by the gluon propagator in QCD,
at least in the gauges considered.

\vspace{0.5cm}
{\bf 2.} To evaluate the propagator we utilize
the Zwanziger local field theory
of electrically and magnetically charged particles~\cite{Zw71}.  The
theory contains two vectors potentials, namely the
gauge field $A_\mu(x)$ and the dual
gauge field $B_\mu(x)$, which interact covariantly with electric and
magnetic currents, respectively. The corresponding
Lagrangian is:
\beqn
\cL = \cL_{Zw}(A,B)+ i \, e \, j^e_\mu (x) \, A_\mu (x)
             + i \, g \, j^m_\mu (x) \, B_\mu (x)\,,
\label{ZwInteraction}
\eeqn
where $e$ ($g$) stands for the electric (magnetic) charge, and $j^e_\mu$
($j^e_\mu$) is the electric (magnetic) external current. The
Zwanziger Lagrangian $\cL_{Zw}$ is given by the following
equation~\cite{Zw71}:
\beqn
\cL_{Zw}(A,B) & = & \frac{1}{2}(n\cdot[\diff\wedge A])^2
+ \frac{i}{2}(n\cdot[\diff\wedge A])(n\cdot\dual{[\diff\wedge B]})
\nonumber\\
& & + \frac{1}{2}(n\cdot[\diff\wedge B])^2
- \frac{i}{2}(n\cdot[\diff\wedge B])(n\cdot\dual{[\diff\wedge A]})\}\,,
\label{ZwLagrangian}
\eeqn
where we have used the standard notations:
\beqn
[A\wedge B]_{\mu\nu} = A_\mu B_\nu - A_\nu B_\mu\,, \quad
(n \cdot [A\wedge B])_\mu = n_\nu (A\wedge B)_{\nu\mu}\,,
\quad \dual{(G)}_{\mu\nu} =
\frac{1}{2} \varepsilon_{\mu\nu\lambda\rho} G_{\lambda\rho}\,.
\eeqn
Despite the theory~\eq{ZwInteraction} apparently
contains two gauge fields, $A_\mu$
and $B_\mu$, there is only one physical particle (a massless photon) in
the spectrum~\cite{Zw71,BaRuSc75}. The Lagrangian
\eq{ZwInteraction} depends on an arbitrary constant unit vector
$n_\mu$, $n^2_\mu = 1$, while physical observables are insensitive to
the direction of $n$ provided the Dirac quantization condition,
$e \cdot g = 2 \pi m$, $m \in \Z$, is
satisfied~\cite{Zw71}.

\vspace{0.5cm} {\bf 3.} Within the dual superconductor
approach~\cite{tHMa76} a pure $SU(2)$ gauge theory is regarded as a
theory of dynamical abelian monopoles which are required to be
condensed in the color confining phase. According to
eq.\eq{ZwInteraction} the theory of monopoles interacting with an
external electric source $j^e$ (quark current) is described by the
following partition function~\cite{MaSu89}:
\beqn
\cZ[j^e] & = & \int \dD A \dD B \dD \Phi
\, \exp\Bigl\{ - \int \dd^4 x \Bigl(
\cL_{Zw}(A,B) + \frac{1}{2} {|(\partial_\mu + i g B_\mu) \Phi|}^2
\nonumber\\
& &  + \lambda {({|\Phi|}^2 - \eta^2)}^2 - i e j^e_\mu
A_\mu\Bigr)\Bigr\}\, .
\label{GeneratingFunction}
\eeqn

Here $\Phi$ is the field of condensed monopoles, the constants
$\lambda,\eta^2$ are positive and we choose to work in the Euclidean
space. In this approach, the abelian field $A_\mu$ corresponds to the
diagonal component $A^3_\mu$ of the $SU(2)$ gauge field $A^a_\mu$ in a
certain abelian gauge. According to the abelian dominance
phenomenon~\cite{SuYo90} the field $A_\mu$ is responsible for the
infrared properties of the theory. For the sake of simplicity we
consider below the London limit, $\lambda \to \infty$. The final
results remain however qualitatively unchanged if we relax the
condition on the coupling $\lambda$.

The spectrum of the
model~\eq{GeneratingFunction} contains a string--like topological
excitation which carries a quantized electric flux. This string is the
dual analogue of the Abrikosov--Nielsen--Olesen (ANO)
string~\cite{ANO} in the Abelian Higgs model. Stretched between
quarks and anti-quarks the string leads to the color
confinement. The confinement phenomenon exists in the theory already
at the classical level, the linear string contribution to the quark
potential is dominant at large distances~\cite{MaSu89,Sug2,GuPoZa98-1}
and survives at small quark--anti-quark separations~\cite{GuPoZa98-2}.

\vspace{0.5cm}
{\bf 4.} The gauge field propagator is defined as:
\beqn
D_{\mu\nu} (x,y) \equiv <A_\mu (x) A_\nu
(y)> = - \frac{1}{e^2} \, \frac{\delta^2}{\delta
j^e_\mu (x) \, \delta
j^e_\nu (y)} \frac{\cZ[j^e]}{\cZ[0]} {\Biggl|}_{j^e = 0}\, ,
\label{PropagatorDefinition}
\eeqn
and is an analogue of the gluon propagator in gluodynamics.

To specify the propagator completely
we choose the axial gauge $n_\mu
A_\mu = 0$ in expressions
(\ref{GeneratingFunction},\ref{PropagatorDefinition}).
Moreover, using the
methods of Refs.~\cite{BKTtoHiggs} we obtain readily a string
representation for~$\cZ[j^e]$:
\beqn
& & \cZ[j^e] \propto \oint\limits_{\partial \Sigma = 0} \dD
\Sigma \, \exp\Bigl\{ - \int \dd^4 x  \int \dd^4 y \Bigl[
\frac{2 \pi^2}{e^2} j^2_\mu(x) D_{M_B}(x-y) \cdot X_{\mu\nu}
j^e_\nu (y) \nonumber\\
& & + 4 \eta^2 \pi^2 j^e_\mu (x) D_{M_B}(x-y) \cdot {(\partial
\cdot n)}^{-1} n_\nu \Sigma_{\mu\nu}  (y) +
\eta^2 \pi^2 \Sigma_{\mu\nu} (x) D_{M_B}(x - y) \Sigma_{\mu\nu} (y)
\Bigr] \Bigr\} \, ,
\label{StrRepr}
\eeqn
where $D_{M_B}$ is the propagator of a massive scalar
field $( - \partial^2_\mu + M^2_B) D_{M_B}(x) = \delta^{(4)} (x)$,
$M_B = e \eta$ is the mass of the dual gauge boson $B_\mu$, and the
integration in eq.\eq{StrRepr} is performed over all closed
world sheets of the ANO strings (see
Refs.~\cite{BKTtoHiggs}).
Moreover, $X_{\mu\nu}$ is a differential operator
which in the momentum representation has the form:
\beqn
X_{\mu\nu} (p) & = &
\delta_{\mu\nu} - \frac{1}{(p \cdot n)} (p_\mu n_\nu + p_\nu n_\mu) +
\frac{p_\mu p_\nu}{{(p \cdot n)}^2} + \frac{M^2_B}{{(p \cdot n)}^2}
(\delta_{\mu\nu} - n_\mu n_\nu)\,. \label{X}
\eeqn

Substituting \eq{StrRepr} in \eq{PropagatorDefinition} we get the
following expression for the propagator:
\beqn D^{\mathrm{axial}}_{\mu \nu}(p) & = & \frac{1}{p^2 + M^2_B}
\cdot X_{\mu\nu} (p) + D^{\mathrm{str}}_{\mu \nu}(p)\,,
\label{PropagatorMomentum}\\
D^{\mathrm{str}}_{\mu \nu}(p) & = & -
\frac{\eta^4 e^2}{p^2 + M^2_B}\cdot \frac{1}{(p \cdot n)}
\cdot \int \frac{\dd^4 k}{{(2\pi)}^2} \frac{1}{k^2 +
M^2_B} \frac{n_\alpha n_\beta}{(k \cdot n)} \cdot {<\Sigma_{\mu\alpha}
(p) \Sigma_{\nu\beta} (-k)>}_\Sigma\,,
\label{PropagatorString}
\eeqn
Note that the expression~\eq{PropagatorMomentum} for the gluon propagator is
exact in the London limit (there are no loop contributions to
eq.\eq{PropagatorMomentum})!

The first term in the propagator is already known from
Ref.~\cite{BaRuSc75} while the second term describes interaction of
the gauge boson with the ANO strings. Moreover the string--string correlation
function in \eq{PropagatorString} is defined as:
\beqn
{<{\cO}>}_\Sigma
& = & \frac{1}{\cZ^{\mathrm{str}}} \,
\oint\limits_{\partial \Sigma = 0} \dD
\Sigma \, e^{- S_{\mathrm{str}}(\Sigma)}\, \cO \,, \quad
\cZ_{\mathrm{str}} = \oint\limits_{\partial \Sigma = 0} \dD
\Sigma \, e^{- S_{\mathrm{str}}(\Sigma)}\,,\\
S_{\mathrm{str}}(\Sigma) & = & \pi^2 \eta^2 \int \dd^4 x \int \dd^4 y
\, \Sigma_{\mu\nu}(x) D_{M_B} (x-y) \Sigma_{\mu\nu}(y)\,.
\label{StrAction}
\eeqn
The expression \eq{StrAction} is the action for the ANO strings in
the Abelian Higgs model in the London limit~\cite{BKTtoHiggs}.

An exact expression for the string--string correlator is unknown.
Taking into account the closeness condition,
$\partial_\mu \Sigma_{\mu\nu} (x) = 0$, one can
parametrize this correlator as follows:
\beqn
{<\Sigma_{\mu\alpha}(x) \Sigma_{\nu\beta} (y)>}_\Sigma
= \varepsilon_{\mu\alpha\xi\rho} \,
\varepsilon_{\nu\beta\zeta\rho}\,
\partial_\xi \partial_\zeta
\, {{\tilde D}}^\Sigma \Bigl({(x - y)}^2\Bigr)\,,
\label{StringCoordinate}
\eeqn
or, in the momentum representation:
\beqn
{<\Sigma_{\mu\alpha} (k) \Sigma_{\nu\beta}
(-p)>}_\Sigma & = & {(2 \pi e)}^2 \delta(k-p) \cdot
\varepsilon_{\mu\alpha\xi\rho} \,
\varepsilon_{\nu\beta\zeta\rho}\,
p_\xi p_\zeta\, {{D}}^\Sigma (p^2)\,,
\label{StringMomentum}
\eeqn
where ${{D}}^\Sigma (p^2) = {{D}}^\Sigma (p^2; e, \eta)$ is a scalar
function, which is related to the function ${{\tilde D}}^\Sigma$ in
eq.\eq{StringCoordinate} via the Fourier transform.

Substituting the string correlation function \eq{StringMomentum}
into eq. \eq{PropagatorString} we get:
\beqn
D^{\mathrm{str}}_{\mu \nu}(p) & = & \frac{M_B^4
}{{(p^2 + M^2_B)}^2} \, D^\Sigma (p^2) \, \Bigl[
\Bigl(\delta_{\mu\nu} - \frac{p_\mu n_\nu
+ p_\nu n_\mu}{{(p \cdot n)}}
+  \frac{p_\mu p_\nu}{{(p \cdot n)}^2}\Bigr) \nonumber\\
& & - \frac{p^2}{{(p \cdot n)}^2}
\cdot (\delta_{\mu\nu} - n_\mu n_\nu)
\Bigr]\,.
\eeqn
Thus, the gluon propagator \eq{PropagatorMomentum}
in the axial gauge has the form:
\beqn
D^{\mathrm{axial}}_{\mu \nu}(p) & = &
\frac{1}{p^2 + M^2_B} \, \Bigl[
\Bigl(\delta_{\mu\nu} +  \frac{p_\mu n_\nu
+ p_\nu n_\mu}{{(p \cdot n)}}
-  \frac{p_\mu p_\nu}{{(p \cdot n)}^2}\Bigr) \cdot
\Bigl( 1 + \frac{M^4_B D^\Sigma (p^2)}{p^2 + M^2_B}\Bigr)
\nonumber\\
& & + \frac{1}{{(p \cdot n)}^2} \cdot (\delta_{\mu\nu} - n_\mu n_\nu)
\cdot M^2_B \, \Bigl( 1 - M^2_B
\frac{p^2 \, D^\Sigma (p^2)}{p^2 +M^2_B}\Bigr)
\Bigr]\,.
\label{prop}
\eeqn

The propagator has singularities not only in
the $p^2$ plane but in the variable $(p\cdot n)$ as well.
Which is not unusual of course
because $n_{\mu}$-dependence enters through the gauge fixing.
What is unusual, however, is that the
singular in $(p\cdot n)$ terms appear not only in the longitudinal
structures but in front of $(\delta_{\mu\nu}-n_{\mu}n_{\nu})$
as well. While the former singularities do not contribute to the
physical amplitudes and can be removed by changing the gauge, the
same is not true in the latter case. The singular factors
in front of the $(\delta_{\mu\nu}-n_{\mu}n_{\nu})$ structure
appear due to the Dirac strings which are present in the Zwanziger
Lagrangian \eq{ZwLagrangian}. The issue of these singularities is
finally settled through observation that the relation between
physical observables and the propagator becomes more subtle than,
say, in QED (see Ref.~\cite{GuPoZa98-1}). Here we are concerned
primarily with formal
properties of the propagator itself.

A general parametrization of the gauge boson
propagator in the axial gauge is ~\cite{BaBaZa88}:
\beqn
D^{\mathrm{axial}}_{\mu\nu} = \Bigl(\delta_{\mu\nu} +
\frac{p_\mu n_\nu + p_\mu n_\nu}{{(p \cdot n)}} - \frac{p_\mu
p_\nu}{{(p \cdot n)}^2}\Bigr) \cdot F(p^2) - \frac{1}{{(p \cdot n)}^2}
\cdot (\delta_{\mu\nu} - n_\mu n_\nu) \cdot G(p^2)\,,
\label{baker}
\eeqn
where $F(p^2)$ and $G(p^2)$ are the
coefficient functions which characterize the
vacuum of the theory. A comparison of eq. \eq{prop} with
eq. \eq{baker} gives the following relation between these coefficient
functions and the string correlation function $D^\Sigma$ introduced above:
\beqn
F(p^2) & = & \frac{1}{p^2 + M^2_B} \, \Bigl( 1 +
\frac{M^4_B D^\Sigma
(p^2)}{p^2 + M^2_B}\Bigr)\,; \label{f}\\
G(p^2) & = & - \frac{M^2_B}{p^2 + M^2_B}
\cdot \Bigl( 1 - M^2_B
\frac{p^2 \, D^\Sigma (p^2)}{p^2 + M^2_B}\Bigr)\,.
\label{g}
\eeqn

Note that the expressions \eq{f} and \eq{g} are non--perturbative even
without taking into account the string contributions. Indeed, removing
for a moment the
string contributions from these relations we get:
\beqn
F^{\mathrm{no-string}}(p^2) = \frac{1}{p^2 + M^2_B}\,,\quad
G^{\mathrm{no-string}}(p^2) = - \frac{M^2_B}{p^2 + M^2_B}\,,
\label{G:no-string}
\eeqn
while for the perturbative vacuum one has~\cite{BaBaZa88}:
\beqn
F^{\mathrm{pert}}(p^2) = \frac{1}{p^2}\,,\quad
G^{\mathrm{pert}}(p^2) = 0\,.
\eeqn
Equations \eq{G:no-string} account for the same graphs as in the
standard Higgs mechanism, {\it i.e.} for insertions of the scalar
field condensate.

\begin{figure*}[!tbh]
\vspace{0.5cm}
\centerline{\epsfxsize=14.1cm\epsffile{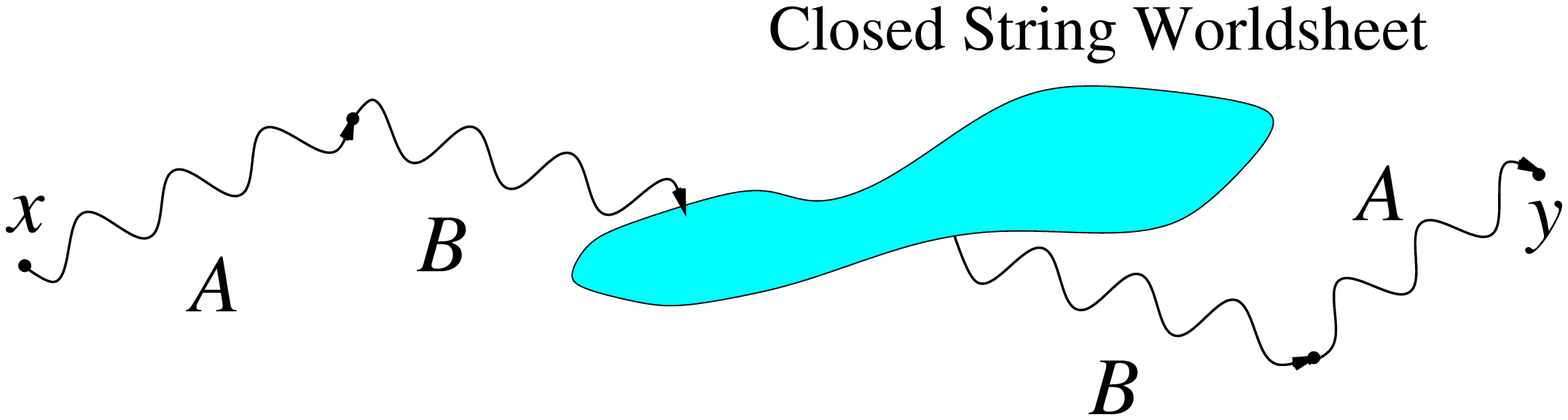}}
\vspace{0.5cm}
\caption{String contribution \eq{PropagatorString} to gluon
propagator \eq{PropagatorMomentum}.}
\label{FigOne}
\end{figure*}

\vspace{0.5cm} {\bf 5.}
Although the expression \eq{prop} for the gluon propagator is exact, it
contains an unknown function $D^\Sigma (p^2)$. The infrared
behaviour of this function
however can be qualitatively understood on physical
grounds. Indeed, consider the string part \eq{PropagatorString} of the gluon
propagator \eq{PropagatorMomentum}. This part corresponds to the
process, schematically shown in Figure~\ref{FigOne}: the quark emits
a gauge boson $A$ which transforms to a dual gauge boson $B$ via
coupling $<A \cdot B>$ which exists in the Zwanziger Lagrangian
\eq{ZwLagrangian}. The dual gauge boson transforms back to the gauge
boson after scattering on a closed ANO string world sheet
$\Sigma$. The intermediate string state is described by the function
$D^\Sigma(p^2)$ and this state can be considered as a glueball state
with the photon quantum numbers $1^-$.

The behaviour of the function $D^\Sigma(p^2)$ in the infrared region,
$p \to 0$, can be estimated as follows:
\beqn
D^\Sigma(p^2) = \frac{C}{p^2 + M_{\mathrm{gl}}^2} + \dots\,,
\label{Dp}
\eeqn
where $C$ is a dimensionless parameter and $M_{\mathrm{gl}}$ is the
mass of the lowest $1^-$ glueball. The dots denote the
contributions of heavier states.
Thus, according to eqs. (\ref{Dp},\ref{f},\ref{g}) the
coefficient functions $F$ and $G$
in the infrared limit, $p \to 0$, become:
\beqn
\begin{array}{ccl}
F^{\mathrm{infr}}(p^2) & = &
\frac{1}{M^2_B} + \frac{C}{M^2_{\mathrm{gl}}}
+ O(p^2)\,,\\
G^{\mathrm{infr}}(p^2) & = & -1 + O(p^2)\,,
\end{array}
\label{infr}
\eeqn

Note that neither double, $1 \slash {(p^2)}^2$, nor ordinary,
$1\slash p^2$, pole
are exhibited by the gluon propagator (\ref{baker},\ref{infr})
in the infrared limit. There are
gauge dependent $1/(p\cdot n)$ singularities which reflect
the presence of Dirac strings. Therefore our gauge boson propagator
does not vanish in the infrared limit either.
In view of this, the vanishing of the propagator at $p\to 0$
predicted in \cite{Zw91} seems to be specific for
the special gauge choice in QCD which leads to a nontrivial
Fundamental Modular Region.

Our demonstration of non-existence of the double pole in the propagator is
based on the infrared behaviour \eq{Dp} of the string--string
correlation function $D^\Sigma$. Suppose for a moment that our
prescription for $D^\Sigma$ is not correct and the double pole exists
in the propagator.
We would conclude then from
eq. \eq{f} that the string
correlation function has a double pole at $p^2=0$.  It
would imply, in turn,
long range correlations of the string world sheets
and, as a result, the absence of the mass gap and quark confinement!
Indeed, if the strings are long-range correlated then the
quarks associated with the string ends would
also be long--range propagating objects.
Thus, the absence of a double pole in the gluon propagator is a
direct consequence of confinement, at least in the dual
superconductor model of infrared QCD considered here.

It is worth emphasizing that our qualitative results do not depend on the
gauge choice. While up to now we discussed the axial gauge,
$n_\mu A_\mu = 0$,
the choice of the Landau gauge, $\partial_\mu A_\mu = 0$, leads to a
mere shift of the operator \eq{X} which enters the definition of the
propagator~\eq{PropagatorMomentum}:
$$X^{\mathrm{Landau}}_{\mu\nu} (p) = X_{\mu\nu} (p) +
\frac{p^2 + M^2_B}{p^2\, (p \cdot n)} (p_\mu n_\nu + p_\nu n_\mu) -
\frac{p^2 + M^2_B}{(p^2)^2\, (p\cdot n)^2} ((p \cdot n)^2 + p^2) \cdot
p_\mu p_\nu\,.$$
Due to this shift an ordinary pole $1 \slash p^2$
appears in a longitudinal term which is irrelevant, however, for
gauge-invariant observables.

To summarize, the gauge boson propagator in the model considered
does not display either a double pole in $p^2$ or vanishing in
the limit $p^2\to 0$, the phenomena thought to be a signature for
the quark confinement \cite{BaBaZa88, Zw91}.
It is rather the stringy singularities in the $(p\cdot n)$ variable
which signal the confinement. The other closest infrared
singularities present in the gauge boson propagator correspond to the
mass of the gauge boson and the lowest $1^-$  glueball mass. Moreover
in our case the confining potential is not related to the infrared
behaviour of the gauge boson propagator in the conventional way:
$\tilde{V}(r) = \int\, e^{ipr} D_{00}(p) \dd ^3 p$. Indeed, $D_{00}$
contains $1/(p\cdot n)$ singularities (see eq.\eq{prop}) related to
the Dirac string, and thus $\tilde{V}(r)$ is gauge dependent. The
right way to obtain the gauge independent expression for the
potential $V(r)$ is to evaluate the expectation value of the Wilson
loop for the field $A_\mu$ in the string representation of the model.
In the string representation the $1/(p\cdot n)$
singularities in the propagator play the crucial role: they transform
into the confining ANO strings. The details of this calculation are given
in Ref.\cite{GuPoZa98-1}.

\vspace{0.5cm} {\bf 6.}
M.N.Ch. and M.I.P. feel much obliged for the kind hospitality
extended to them at the
Max-Planck-Institute for Physics in Munich where
this work has been started. M.I.P wishes to express his gratitude to the
members of the Department of Physics of the Kanazawa University where
this work has been completed. This work was partially supported by the
grants INTAS-96-370, INTAS-RFBR-95-0681, RFBR-96-02-17230a and
RFBR-96-15-96740. The work of M.N.Ch. was supported by the INTAS Grant
96-0457 within the research program of the International Center for
Fundamental Physics in Moscow. M.I.P. was supported by JSPS (Japan Society
for the Promotion of Science) Fellowship Program.


\begin{thebibliography}{99}

\bibitem{tHMa76} G.~{'t Hooft}, "High Energy Physics",
{\rm Ed. by A.~Zichichi, Editrice Compositori, Bolognia}, 1976;\\
S.~Mandelstam, {\it Phys.~Rept.} {\bf 23C} (1976) 245.

\bibitem{tH}
G.~'t Hooft, {\it Nucl. Phys.} {\bf B190} (1981) 455.

\bibitem{Reviews}
T.~Suzuki, {\it Nucl. Phys.} {\bf B} {\it (Proc.~Suppl.)} {\bf 30}
(1993) 176;\\
M.I.~Polikarpov, {\it Nucl. Phys.} {\bf B} {\it (Proc.~Suppl.)} {\bf
53} (1997) 134;\\
M.N.~Chernodub and M.I.~Polikarpov, in {\it "Confinement,
Duality and Non-perturbative Aspects of QCD"}, p.387, Ed. by Pierre van Baal,
Plenum Press, 1998; {\tt hep-th/9710205};\\
G.S.~Bali, {\it preprint HUB-EP-98-57, Jan. 1998},
{\tt hep-ph/9809351}.

\bibitem{West}
G.B.~West, {\it Phys.~Lett.} {\bf 115B} (1982) 468.

\bibitem{BaBaZa88} M.~Baker, J.S.~Ball and F.~Zachariasen, {\it
Phys.~Rev.} {\bf D37} (1988) 1036; {\it Erratum-ibid.} {\bf D37}
(1988) 3785.

\bibitem{Zw91}
V.N.~Gribov, {\it Nucl.~Phys.} {\bf B139} (1978) 1;\\
D.~Zwanziger, {\it Nucl.~Phys.} {\bf B364} (1991) 127.

\bibitem{alagribov}
U.~H\"abel {\it et al.}, {\it Z.Phys.} {\bf A 336} (1990) 423;\\
L.~von Smekal, A.~Hauck and R.~Alkofer,
{\it Phys. Rev. Lett.} {\bf 79} (1997) 3591; {\it Ann. Phys.}
{\bf 267} (1998) 1;\\
D.~Atkinson and J.C.R.~Bloch, {\it Mod. Phys. Lett}
{\bf A13} (1998) {1055}.

\bibitem{nogribov} M.~Baker, J.S.~Ball and F.~Zachariasen, {\it Nucl.
Phys.}  {\bf B186} (1981) 531;\\
N.~Brown and M.R.~Pennington, {\it Phys. Rev.}
{\bf D39} (1989) 2723;\\
A.I.~Alekseev, {\it Phys. Lett.}
{\bf B344} (1995) 325;\\
F.T.~Hawes, P.~Maris, C.D.~Roberts, {\it Phys. Lett.}
{\bf B440} (1998) 353.

\bibitem{ZwNumerical}
C.~Bernard, C.~Parrinello and A.~Soni, {\it Phys.~Rev.}
{\bf D49} (1994) 1585;\\
H.~Nakajima and S.~Furui, {\tt hep-lat/9809081}, {\tt hep-lat/9809078};\\
A.~Cucchieri, {\it Phys.~Lett.} {\bf B422} (1998) 233;\\
D.~B.~Leinweber {\it et al.};
{\it preprint ADP-98-72/T339}, {\tt hep-lat/9811027};\\
J.P.~Ma, {\it preprint AS-ITP-99-07} {\tt hep-lat/9903009}.

\bibitem{MaSu89} S.~Maedan and T.~Suzuki, {\it Prog.~Theor.~Phys.}
{\bf 81} (1989) 229.

\bibitem{effmod}
S.~Kato, S.~Kitahara, N.~Nakamura and T.~Suzuki,
{\it Nucl.~Phys.} {\bf B520} (1998) 323;\\
M.N.~Chernodub {\it et al.},
{\it preprint KANAZAWA-98-19}, {\tt hep-lat/9902013}.

\bibitem{Zw71}
D. Zwanziger, {\it Phys. Rev.} {\bf D3} (1971) 343;\\
R.A. Brandt, F. Neri, and D. Zwanziger, {\it Phys. Rev.} {\bf D19}
(1979) 1153;\\
M.~Blagoevi\'{c} and Senjanovi\'{c}, {\it Phys. Rep.}
{\bf 157} (1988) 233.

\bibitem{BaRuSc75}
A.P. Balachandran, H. Rupertsberger, and J. Schechter,
{\it Phys. Rev.} {\bf D11} (1975) 2260;\\
T.~Suzuki,  {\it Prog.~Theor.~Phys.} {\bf 81} (1989) 752.

\bibitem{SuYo90}
Z.F.~Ezawa and A.~Iwazaki,
{\it Phys.Rev.} {\bf D25} (1982) 2681;\\
T.~Suzuki and I.~Yotsuyanagi, {\it Phys.~Rev.}, {\bf
D42} (1990) 4257;\\
G.S.~Bali {\it et.~al.}, {\it Phys.~Rev.} {\bf D54} (1996) 2863.

\bibitem{ANO}
A.A.~Abrikosov, {\it JETP} {\bf 32} (1957) 1442;\\
H.B.~Nielsen and P.~Olesen, {\it Nucl.~Phys.} {\bf B61} (1973) 45.

\bibitem{Sug2} H.~Suganuma, S.~Sasaki and H. Toki, {\it Nucl.~Phys.}
{\bf B435} (1995) 207;\\
S.~Sasaki, H.~Suganuma and H.~Toki, {\it Phys.Lett.}
{\bf B387} (1996) 145.

\bibitem{GuPoZa98-1} F.V.~Gubarev, M.I.~Polikarpov and V.I.~Zakharov,
{\it Phys.~Lett.} {\bf B438} (1998) 147.

\bibitem{GuPoZa98-2} F.V.~Gubarev, M.I.~Polikarpov and
V.I.~Zakharov, {\it preprint \it ITEP-TH-73-98},\\ {\tt
hep-th/9812030}.

\bibitem{BKTtoHiggs}
P.~Orland, {\it Nucl. Phys.} {\bf B428}, (1994) 221;\\
M.~Sato and S.~Yahikozawa,
{\it Nucl. Phys.} {\bf B436} (1995) 100;\\
E.T. Akhmedov {\it et al.},
{\it Phys. Rev.} {\bf D53} (1996) 2087;\\
M.I.~Polikarpov, U.-J.~Wiese and M.A.~Zubkov, {\it Phys.~Lett.}, {\bf
309B} (1993) 133.

\end{thebibliography}
\end{document}